\documentclass[useAMS,usenatbib,a4]{mn2e}

\usepackage{graphicx} 
\usepackage{amsmath} 
\usepackage{amssymb} 
\usepackage{rotating}   
\usepackage{multirow}
\usepackage{units}
\usepackage{subfigure}
\def\deg{^{\circ}}
\def\P3hat{{\mathaccent 94 P}_3}

\def\eg{{\it e.g.}}
\def\ie{{\it i.e.}}

\newcount\notenumber
\def\clearnotenumber{\notenumber=0}
\def\note{\advance\notenumber by1 \footnote{$^{\the\notenumber}$}}
\clearnotenumber

\title[Pulsar B0943+10: An 8-Hour GMRT Observation]{The topology and polarization of subbeams associated\\
with drifting subpulse emission of pulsar B0943+10 --- \\VI. Analysis of an 8-hour GMRT observation}

\author[Isaac Backus, Dipanjan Mitra, \& Joanna Rankin] 
{Isaac Backus$^{1,2}$, Dipanjan Mitra$^{2}$ \& Joanna M. Rankin$^{1}$ \\ 
$^1$Physics Department, University of Vermont, Burlington, VT 05405 USA\thanks{Isaac.Backus@uvm.edu; Joanna.Rankin@uvm.edu} \\
$^2$National Centre for Radio Astrophysics, Ganeshkhind, Pune 411 007 India\thanks{dmitra@ncra.tifr.res.in; } 
}
\date{Unreleased}

\pagerange{\pageref{firstpage}--\pageref{lastpage}} \pubyear{2004}

\def\LaTeX{L\kern-.36em\raise.3ex\hbox{a}\kern-.15em
    T\kern-.1667em\lower.7ex\hbox{E}\kern-.125emX}

\begin{document}

\label{firstpage}

\maketitle

\begin{abstract}
We report analysis of an 8~hr observation of PSR B0943+10 at 325~MHz performed at the Giant Metrewave Radio Telescope (GMRT) in India.  B0943+10 is well known for displaying regular sub-pulse drifting and two emission modes.  We investigate the modal behavior of B0943+10.  By reconstructing an entire `B' mode from two consecutive `B' modes, we estimate that the pulsar spends roughly 7.5~hrs in the `B' mode and about 2.2~hrs in the `Q' mode, on average.  Although the pulsar can switch modes within one pulse, the sub-pulse drift rate changes with a characteristic time of 1.2~hrs over the course of a `B' mode.  Under the subbeam carousel model we find the drift-rate changes are produced by a 10\% increase in the average number of subbeams and a 16\% increase in the carousel circulation time.  We speculate that under the partially screened gap model the increase in circulation time should be related to a small increase in the neutron star surface temperature.
\end{abstract}


\section*{I. Introduction} 
\label{sec:I}
Pulsar B0943+10 is well known as a paradigm example of moding and subpulse drifting in pulsars.  In the `B'urst mode, the pulsar displays strong, regular subpulse drifting, while in the `Q'uiescent mode the drifting disappears and only hints of a modulation feature have been found [\cite{PaperIV}].  During the `Q' mode, B0943+10 grows dimmer and a precursor component (hereafter:~PC) appears whose peak precedes the peak of the main pulse (hereafter:~MP) by some 50$\deg$, as discovered in \cite{Backus}.

The papers in this series\footnote{
The papers in the \textit{Topology and Polarization} series are: 
\cite{PaperI} [hereafter: \textit{Paper~I}]; 
\cite{PaperII} [\textit{Paper~II}]; 
\cite{PaperIII} [\textit{Paper~III}]; 
\cite{PaperIV} [\textit{Paper~IV}]; 
and \cite{PaperV} [\textit{Paper~V}].}
 have investigated at length the dynamic behaviors of B0943+10.  Letting $P_1$ be the pulsar period, a regular modulation feature appearing at $\sim 0.47 / P_1$ (hereafter: $f_{3,obs}$) was found.  In \textit{Paper~I} it was argued that this was a first order alias of the true modulation frequency lying at $\sim 0.53 / P_1$ (hereafter: $f_3$).  \textit{Paper~IV} argued that there are gradual changes in $f_3$ over the course of a `B'-mode interval with the remarkably long time scale of 1.2~hrs.  

This series has investigated whether B0943+10's highly regular drift modulation might be understood on the basis of the subbeam carousel model (hereafter: CM).  It was demonstrated that the subpulse pattern of B0943+10 is produced by a carousel of 20 subbeams rotating with a circulation time $\P3hat$ of some 37 $P_1$.  The projection of the carousel on a plane perpendicular to the magnetic axis is treated as quite circular.

Physically, the carousel is thought to be produced by `spark'-induced columns of relativistic primary plasma directed into the open polar flux tube.  Under ${\bf E}\times{\bf B}$ drift the plasma columns precess around the magnetic axis.  This model was introduced by \cite{RS1975} and has been further developed since [see \cite{Gil2000}].

In this paper we present a nearly 8-hr long observation of B0943+10 carried out using the Giant Metrewave Radio Telescope (GMRT) in India.  This long observation has allowed us to confirm the slow changes of the pulsar's behavior in its `B' mode as discussed in \textit{Papers~IV}~\& \textit{V}.

\S II provides a detailed description of the long GMRT observation as well as of the older Arecibo Observatory (AO) observations used below.  In \S III we investigate the modal evolution of B0943+10 and confirm the gradual variation of $f_3$ during the `B' mode with a time scale of 1.2~hrs.  This result allows confirmation of the slow `B'-mode profile-shape changes with the same time scale (\textit{Paper~IV}~\& \textit{V}).  Overall we found that the `B'- and `Q'-mode intervals had typical durations of some 7.5 and 2.2 hrs, respectively.  \S IV investigates the observed change in the primary modulation frequency $f_3$.  We resolve a decrease in the subpulse spacing ($P_2$) of about 10\% over the course of a `B' mode.  We find that the carousel angular velocity (hereafter:~$\omega_c$ [= $2\pi/\P3hat$]) decreases by nearly 20\%, corresponding to a nearly 20\% increase in $\P3hat$.  Finally \S V provides a discussion of these results.

\section*{II. Observations} 
\label{sec:II} 

\subsection*{1. Observing setup}

The 325-MHz pulse sequences (hereafter: PS) were acquired using the Giant Meterwave Radio Telescope (GMRT) near Pune, India. The GMRT is a multi-element aperture-synthesis telescope \citep{Swarup1991} consisting of 30 antennas distributed over a 25-km diameter area which can be configured as a single dish in both coherent and incoherent array modes of operation. The observations discussed here used the coherent (or more commonly called ``phased array'') mode \citep{Gupta2000,Sirothia2000} in the upper of the two 16-MHz ‘sidebands’. At either frequency right- and left-hand circularly polarized complex voltages arrive at the sampler from each antenna. The voltage signals were subsequently sampled at the Nyquist rate and processed through a digital receiver system consisting of a correlator, the GMRT array combiner (GAC), and a pulsar back-end. In the GAC the signals selected by the user are added in phase and fed to the pulsar backend. The pulsar back-end computes the power levels, which were then recorded at a sampling interval of 0.512 msec.

\subsection*{2. Data reduction, interference mitigation}
In order to put the GMRT observation of B0943+10 into serviceable shape, there were several interference and observational artifacts which had to be addressed.  Due to instrumental effects, the baseline varied on timescales of 10~s or more.  To cope with this, a running mean subtraction was performed.  The on-pulse regions were gated and replaced by random samples from nearby off-pulse regions.  2~s and then 5~s running means were performed accordingly and subtracted from the time series.

Electrical interference at 50~Hz was strong.  To mitigate this, after performing the running mean subtraction, an FFT of about 1 million samples was performed on the dedispersed time series, beginning at the first sample.  In the frequency domain, the real and imaginary components corresponding to the regions around the fundamental frequency and its harmonics were replaced by randomly selected nearby samples.  An inverse FFT was performed and the same process was repeated on the remaining samples in the time series.  This process could not remove all the 50-Hz interference, due to its unstable frequency and spectral leakage.  To account for spectral leakage, large regions around the fundamental frequency and its harmonics would have needed to be removed, but this would have interfered with the pulsar signal and distorted the time series.

Finally, an apparent gradual shift in the phase of the pulsar of about 10$\deg$ during the 8-hr observation was present in the data.  2500-pulse averages were performed and the center of the MP profile was determined for each section of the data.  The period used to bin the data into individual pulses was adjusted over the course of the observation to account for the small changes in phase of the profile.

\subsection*{3. Overview of reduced observation}

\begin{figure}
\begin{center}
\includegraphics[width=80mm,angle=-90]{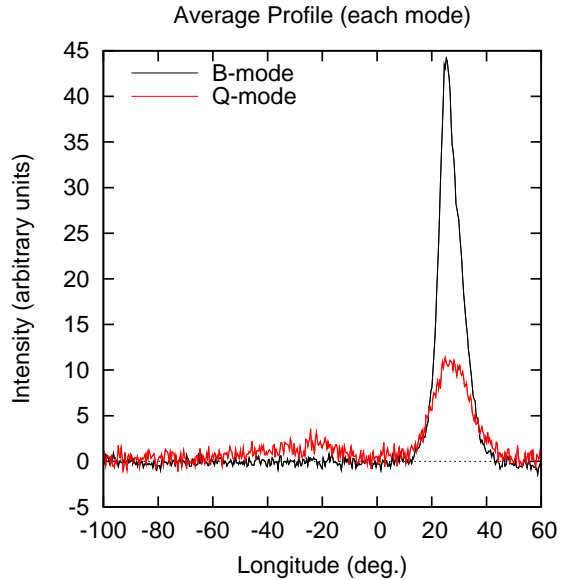}
\caption{Average intensity profiles for both modes.  The `B' mode (in black) is much brighter than the `Q' mode (in red).  The `B' mode profile is taken from several thousand pulses in the first `B' mode apparition (B1).  There is a $\sim5\sigma$ detection of the PC lying $\sim$50$\deg$ before the MP during the `Q' mode, agreeing with \protect\cite{Backus}.  While the MP is dimmer during the `Q' mode, it is significantly wider.} 
\label{fig:profiles}
\end{center}
\end{figure}

B0943+10 is not very bright at 325 MHz, and so the increased sensitivity available at AO made the observations reported in \textit{Papers~I \& IV} significantly higher quality than the long observation presented here.  This prevents certain analyses performed in previous papers, but the great length of this observation opens a new path for inquiry.

We confirm the presence of a precursor (PC) lying 52$\deg$ before the main pulse (MP) observable in the total intensity only during the `Q' mode (see Fig.~\ref{fig:profiles}).  After averaging several thousand `Q' mode pulses, the PC is visible at a peak level of about 5 times the off-pulse rms.  As with all previous observations, the MP is brighter in the `B' mode, and dimmer but wider in the `Q' mode.

\begin{figure*}
\begin{center}

\subfigure[Pulse seqeunce before break (1621 pulses)] {
\includegraphics[width=80mm]{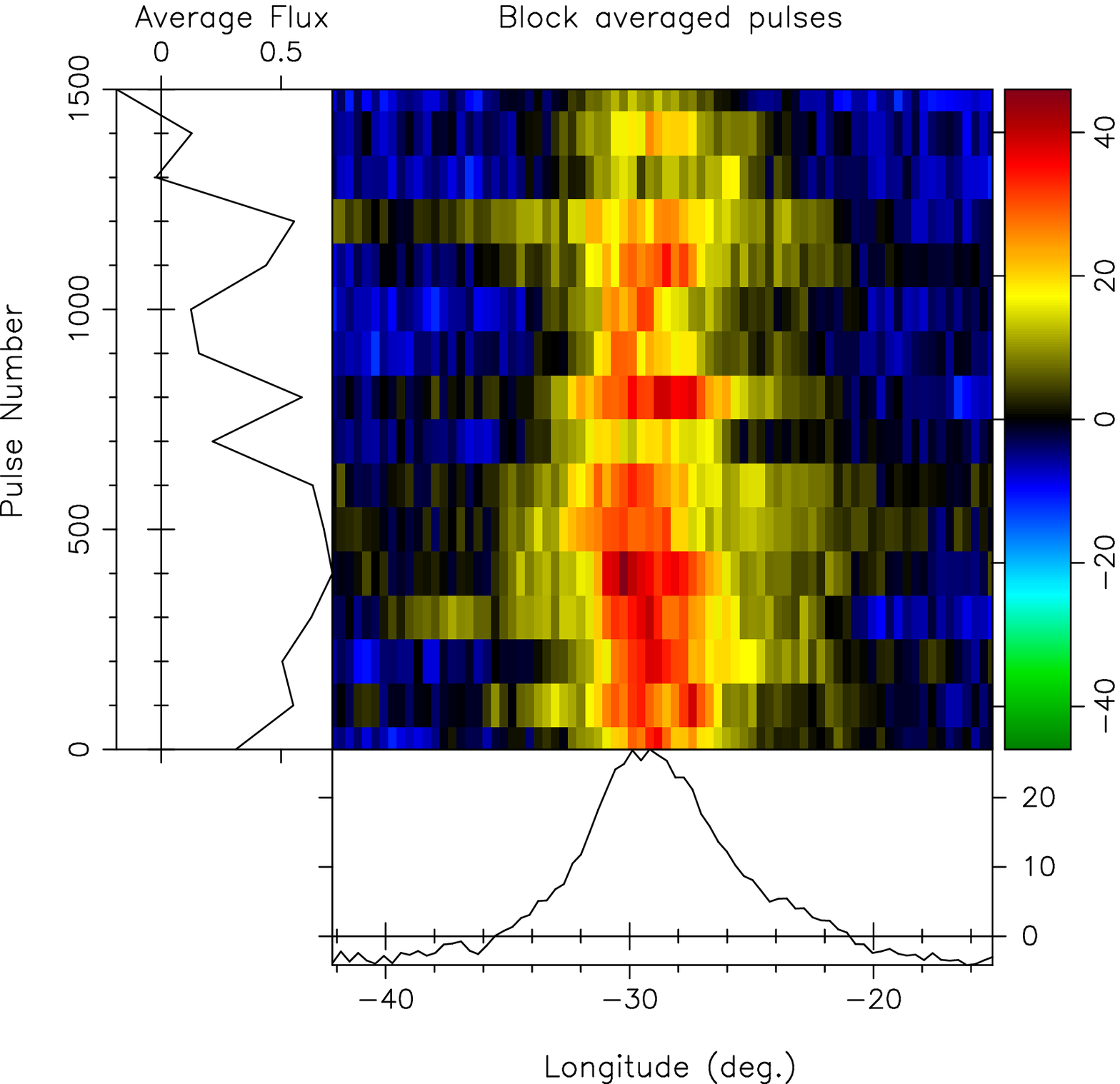}
\label{fig:2nd_obs_ps_a}
}
\subfigure[Pulse sequence after break (23209 pulses)] {
\includegraphics[width=80mm]{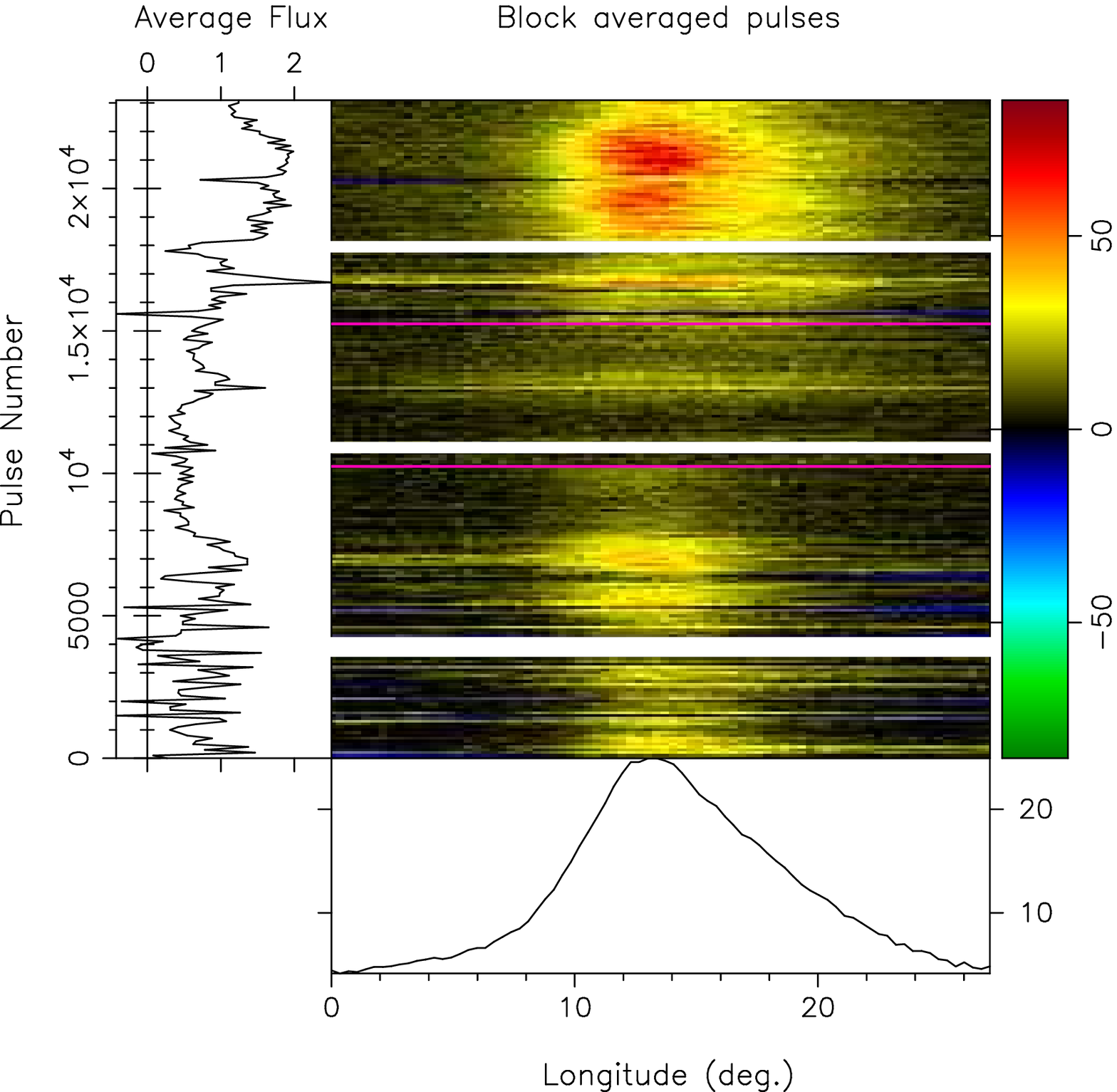}
\label{fig:2nd_obs_ps_b}
}
\caption{Pulse sequences for the January, 2010 GMRT observation at 325 MHz.  (a) shows the first 1621 pulses of the observation, (b) shows the remaining 23209 pulses which were observed following a 13-minute break in the observation.  The horizontal magenta lines in the center panel of (b) mark the beginning and end of the `Q' mode.  The mode changes were found by examining pulse shape changes and the strong modulation features which are only present in the `B' mode.  Due to scintillation and de-phasing of the instrument there are apparent changes in brightness which do not coincide with the actual modal boundaries.  The white lines mark the three occasions where the telescope went off-source to rephase the antennae.  The center panel shows a stack of 100-pulse averages, with time increasing along the +y-axis.  The left panels show the integrated flux for each 100-pulse average and the bottom panel shows the profile averaged over all pulses in the window.  27$\deg$ around the MP are displayed.  The intensity scale and the 0$\deg$ longitude are arbitrary.}
\label{fig:2nd_obs_ps}
\end{center}
\end{figure*}

The most important aspect of this observation is its sheer length.  The observation spans almost 8~hr, with 23,160 pulses available (after truncation due to running mean subtraction).  There were four breaks in the observation.  After 30-min the telescope was stopped for about 13~min in order to recheck the phasing.  Over the following 7~hr there were three $\sim$10~min breaks in the observation spaced more or less evenly during which time the telescope was pointed to a bright nearby source to rephase the antennae.  Figure~\ref{fig:2nd_obs_ps} shows the pulse sequence for this long observation as a stack of 100-pulse averages.  Figure~\ref{fig:2nd_obs_ps_a} shows the 30~min of observation before the long 13-min break.  Figure~\ref{fig:2nd_obs_ps_b} shows the remaining $\sim$7~hr following the break.

The length of the modes in B0943+10 is on the order of 2 or more hours, so it is highly unlikely that an entire mode would be missed due to such breaks, although the boundary between two modes could occur during an observational break.  Luckily, as can be seen in figure~\ref{fig:2nd_obs_ps_b}, the two observed mode changes did not occur during breaks, allowing a good estimation of the position of the modal boundary.

Because of the importance of the long 7-hr stretch after the initial 13-min break, we will follow the convention of setting pulse 0 to the beginning of this stretch.  The modal sequence in this observation is as follows: `B' mode from pulses --2409 to 10270; `Q' mode from pulses 10270 to 15250; then back to `B' mode from pulses 15250 to 23209.  The observational breaks fall at the following pulse intervals:  $-788$ to 0; 3,375 to 4,250; 10,705 to 11,100; 17,775 to 18,175.

\section*{III. Gradual modal evolution}
\label{sec:III}

\textit{Paper~IV} announced the discovery of behaviors associated with the evolution of the modes in B0943+10.  At `B' mode onset, gradual changes in the relative amplitudes of the leading (A1) and trailing (A2) components of the MP were detected, with the ratio of the amplitudes A(2/1) decaying exponentially from 1.75 to about 0.2 (at 327 MHz).  The carousel circulation time (hereafter: $\P3hat$) at `B'-mode onset was found to increase by about 5\% as an exponential of the form $[1-exp(-t/\tau)]$, where $\tau$ is the characterstic time.  We argue for a larger increase of some 16\%.  For both the change in A(2/1) and $\P3hat$, the characteristic time was found to be some 73 minutes, or 1.2 hours.  (see Figs 5 and 8, \textit{Paper~IV})  \textit{Paper~V} found that the fractional linear polarization of the MP increases from 3 to 50\% over the course of a `B' mode.

However, a full justification for these results remained elusive.  Such dynamics can have important implications for understanding the origins of modes in pulsars, as long as it can be shown that these changes occur consistently and can therefore be linked with the evolution of a mode.  \textit{Paper~IV} only had access to observations about 2 hours long at the longest, which introduced several obstacles: an entire mode was never observed from start to finish; only one `B' to `Q' mode transition had been seen (and that observation was only 986 pulses long); and the longest the pulsar had been observed from `B' mode onset onward was some 3900 pulses (1.2 hr).  \textit{Paper~V} only had access to short observations 4 to 15 min in length and determined the dependence of the fractional linear polarization as a function of time after `B' mode onset from the apparent time dependence of $\P3hat$ established in \textit{Paper~IV}.

The decrease of the amplitude ratio A(2/1) and the increase of $\P3hat$ throughout the `B' mode both depended on an imprecise argument involving the splicing together of data from several observations by noting the apparent exponential nature of these changes at `B' mode onset.

We confirm the evolution of the frequency modulation during the course of `B' mode as determined in \textit{Paper~IV}.  To make the timescale of the changes determined in this paper independent of aliasing and model, we examine the frequency of the observed primary modulation feature ($f_{3,obs}$), rather than $\P3hat$ as in \textit{Paper~I--V}.  Under the carousel model, $\P3hat = N/f_{3,true}$ where N is the number of subbeams and $f_{3,true}$ is the un-aliased frequency of the primary modulation feature.  \textit{Paper~I} and \cite{Backus} argued that the observed modulation is a first order alias of the true feature such that $f_{3,true} = 1 - f_{3,obs}$, and that N = 20.  Accordingly, $\P3hat = 20/(1-f_{3,obs})$, so the timescales of the small changes in $\P3hat$ will be reflected in the timescales observed in $f_{3,obs}$ given the gradual exponential behavior of the changes.

\begin{figure*}
\begin{center}

\subfigure[Observed primary modulation frequency ($f_{3,obs}$) for the 2010 GMRT observation of B0943+10.  The pulsar is in the `Q' mode approximately from pulse 10270 to pulse 15250---indicated by the vertical blue stripe and labeled Q1---where no regular modulation is detected.  In the first `B'-mode apparition (B1), $f_{3,obs}$ fluctuates around 0.47 cycles/$P_1$.  At the onset of the second `B' mode (B2), $f_{3,obs}$ increases from $\sim$0.45 cycles/$P_1$ to $\sim$0.47 cycles/$P_1$.  Using an exponential fit (eq.\ref{eq:f3obs}), the characteristic time of this increase is 73 min, or 1.2 hr.] {
\includegraphics[width=80mm]{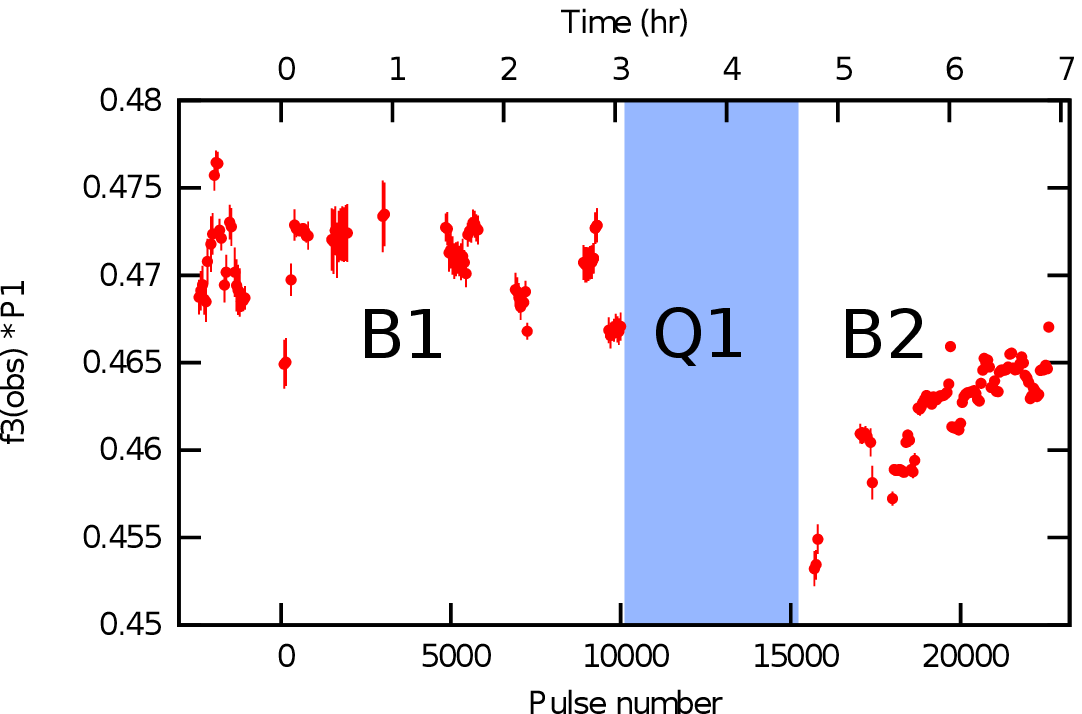}
\label{fig:p3hat_v_time_a}
}
\hspace{5 mm}
\subfigure[$f_{3,obs}$ for the 2010 GMRT observation of B0943+10 as in (a), but with the sections rearranged.  Pulse 0 is set to the beginning of region B2, corresponding to pulse 15250 in (a), and the region B1 is placed to begin at about pulse 11,000.  An exponential was fitted to the data (see eq.\ref{eq:f3obs}) and is plotted as a solid black line.  The ~7.5 hrs from `B' mode onset at the beginning of B2 until `Q' mode onset at the end of B1 represents a good lower limit on the length of `B' modes.  The fit appears to slightly underestimate $f_{3,obs}$ at the beginning of B1 and overestimate it at the end of B1.  This may indicate slight deviation from exponential behavior in $f_{3,obs}$.  However, as is indicated in the caption below, adjacent points carry largely redundant information.  A strong modulation feature at around pulse 12000 which lies around 0.475/$P_1$ may be skewing adjacent points to be farther above the fit line.
] {
\includegraphics[width=84mm]{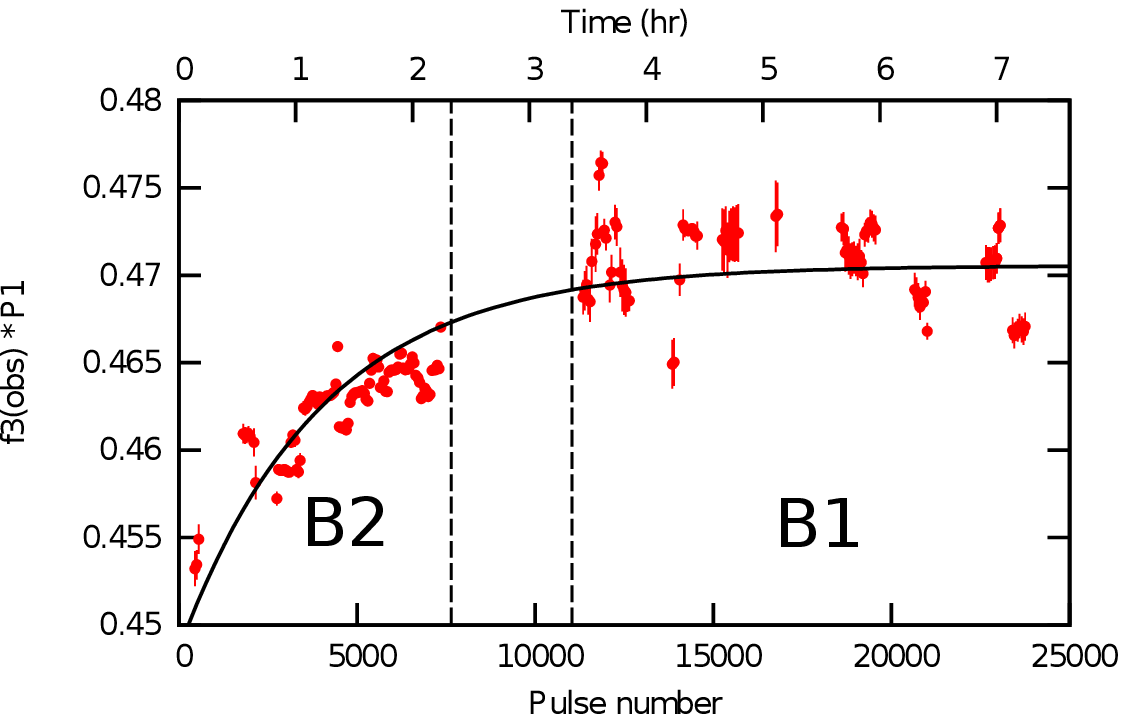}
\label{fig:p3hat_v_time_b}
}
\caption{Observed primary modulation feature ($f_{3,obs}$) as a function of time for the Jan 2010 GMRT observation.  (a) The entire 8-hr observation.  (b) Entire reconstructed `B' mode from the data in (a).  $f_{3,obs}$ was calculated as follows.  512-point LRF's were performed every 50 pulses.  Adjacent LRF's are largely redundant, but can help indicate the stability of a modulation feature.  A sinc function was interpolated around the bin with maximum power in the LRF to find the frequency of the modulation and the error on that measurement.  LRF's where the modulation was poorly defined were discarded.  Due to significantly higher RFI during the first half of the observation, the error bars in B1 are larger than those in B2.}
\label{fig:p3hat_v_time}
\end{center}
\end{figure*}

Figure \ref{fig:p3hat_v_time_a} shows a plot of $f_{3,obs}$ vs pulse number for the long January, 2010 GMRT observation.  $f_{3,obs}$ was determined by performing 512-point Longitudinally Resolved Fluctuation Spectra (hereafter: LRF, see \textit{Paper I}) every 50 pulses.  LRF's with a poorly defined primary modulation feature were discarded.  Each LRF overlaps the next by 462 pulses, so adjacent points are largely redundant, but allow the stability of the feature to be determined.

We confirm the characterstic time.  We find that $\tau = 73 \pm 6$~min ($\sim$1.2 hr) for the increasing $f_{3,obs}$, which is consistent with the findings of \textit{Paper IV}.  Fitting an exponential function to the circulation time as function of pulse number (see Fig.~\ref{fig:p3hat_v_time}) for the second `B'-mode apparition, we find
	\begin{equation}
f_{3,obs}(t) \approx 0.471 - 0.022 e^{-\nicefrac{t}{\tau}} \pm 0.002 /P_1
	\label{eq:f3obs}
	\end{equation}
where $t=0$ at `B'-mode onset.  While we find that $\P3hat$ changes more than was reported in \textit{Paper IV} (see \S IV), eq.(\ref{eq:f3obs}) actually agrees with $f_{3,obs}$ as measured in \textit{Paper IV}.  Significantly, B0943+10 was observed for 7960 pulses after `B'-mode onset, more than twice as long as in any previous observation, and the exponential behavior continues for the entire time.

It should be stressed that using an exponential of the form of eq.(\ref{eq:f3obs}) to describe $f_{3,obs}$ is not physically or theoretically motivated.  There is no model which predicts such a functional form.  The primary purpose of such a fit is to describe the timescale of the changes in $f_{3,obs}$ and the amplitude of the relative changes in $f_{3,obs}$ in a reproducible fashion.

With the $\sim$4 hours of observation before `Q'-mode onset we see very clearly that at the end of a `B' mode, $f_{3,obs}$ has increased to about 5\% higher than at `B' mode onset.  $f_{3,obs}$ is largely constant during the first `B' mode apparation which is to be expected at the end of `B' mode if $f_{3,obs}$ follows the form of eq.(\ref{eq:f3obs}).  It is also observed that $f_{3,obs}$ has decreased 5\% following the `Q' mode, implying that the relation in eq.(\ref{eq:f3obs}) is indeed a regular feature of the `B' mode and is not an illusion due to observations spaced over the course of days or years.

We are now prepared to sketch out the long-term evolution of the modes.  From this long observation it is clear that $f_{3,obs}$ changes exponentially as a function of the form of eq.(\ref{eq:f3obs}) during a single `B' mode.  If we consider this change to be continuous, then we may reconstruct an entire `B' mode from the data within this long GMRT observation.  Fig.~\ref{fig:p3hat_v_time_a} shows $f_{3,obs}$ as a function of time as observed.  Fig.~\ref{fig:p3hat_v_time_b} shows $f_{3,obs}$ as a function of time for such a reconstruction.

To reconstruct a `B' mode from start to finish as in Fig.~\ref{fig:p3hat_v_time_b}, the begining of the second `B' mode apparition (B2) was set to $t=0$.  $f_{3,obs}$ at the beginning of the first `B' mode apparition (B1) is greater than at the end of B2, indicating that B1 comes later in the modal evolution than does B2.  Thus, the first `B' mode (B1) was shifted in time to come after B2.

To determine a reasonable spacing between the end of B2 and the beginning of B1 in the reconstructed sequence, an exponential function of the form of eq.(\ref{eq:f3obs}) was fitted to the $f_{3,obs}$ in both regions for spacings between 0 and 55,000 pulses (enough to be essentially infinite).  The fit is mostly constrained by the more steeply varying values in B2, so the constants in eq.(\ref{eq:f3obs}) are not affected significantly by the spacing ($\Delta$).  The rms of the residuals was found to decay exponentially as $rms = a+be^{-\Delta/\Delta_c}$ where the characteristic spacing $\Delta_c=3250$~pulses.  Thus, the quality of the fit remains nearly constant for $\Delta\geq4000$~pulses.

This sets a lower limit on the length of the `B' mode.  We find the `B' mode should be at least 24,600 pulses (7.5 hr).  We can review t	he plausibility of this as a lower limit, as well as a rough estimate of the actual `B' mode length by considering the fraction of time B0943+10 is observed to be in each mode.  We have 31.2 hrs of observations available composed of 24.0 hr in `B' mode and 7.2 hr in `Q' mode.  The pulsar is therefore observed to be in the `B' mode 77\% of the time and in the `Q' mode the remaining 23\%.  

If the average `B' mode length is 7.5 hrs, then we can expect the average `Q'-mode length to be 2.2 hrs, which seems to agree with observation.  In the GMRT observation, the `Q' mode is 1.5 hrs long; whereas in the only other complete `Q' mode known \citep{Rankin2008}, the duration is almost exactly 1 hr.  In no other observation is an entire `Q' mode seen, but there are two other instances where more than 1.5 hrs of continuous `Q' mode is observed.  That no other complete `Q' modes are observed is unsurprising if `Q' modes average 2.2 hr, given the $\sim$2.5 hr that B0943+10 is visible to AO.

It should be stressed that these timescales are estimates and are not expected to be exact, yet they do provide insight.  Significant deviation from these values for a given mode should be expected.

We may now re-address the findings of \textit{Papers IV} \& \textit{V}.  We confirm that the exponential changes in $f_{3,obs}$ as a function of time persist throughout the `B' mode, thus allowing the estimation of the location of given observation in the `B' mode evolution by measuring its $f_{3,obs}$ (or $\P3hat$ as in \textit{Paper IV} and \textit{V}).  Determining the exponential decay in the ratio of the trailing and leading components of the profile [A(2/1)] was dependent on knowing that $f_{3,obs}$ indeed evolves exponentially with the same timescale found in this paper.  Finally, the increase in fractional linear polarization as a function of location in `B' mode depended on estimating the position of short observations in `B' mode from $f_{3,obs}$.

\begin{figure}
\begin{center}
\includegraphics[width=80mm]{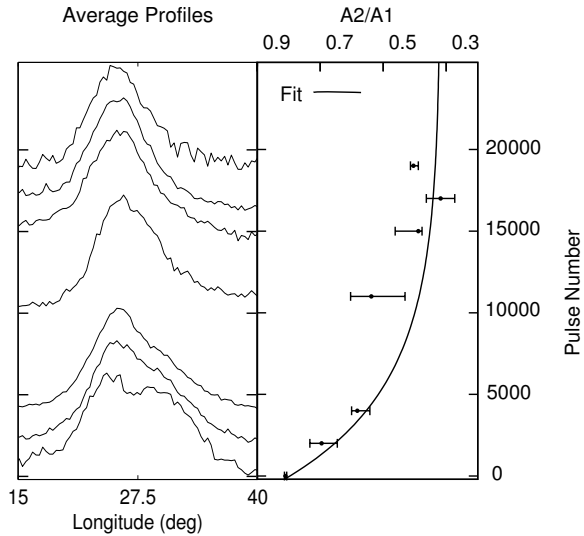}
\caption{\textbf{Left panel:} Normalized average profiles versus pulse number for the reconstructed `B' mode as in Fig.~\ref{fig:p3hat_v_time_a}. Notice that the trailing component is prominent at pulse 0 and decays relative to the leading component as time progresses.  \textbf{Right panel:} Ratio of the amplitudes of the trailing and leading components in the profile (A2/A1).  The fit (solid black line) is an exponential of the form $a+b \exp(-t/\tau)$ where the characteristic time $\tau = 1.7 \pm 1.0 hrs$ agrees with the characteristic time found in \textit{Paper IV}.  The profiles are taken from 2000-pulse averages.  Pulse 0 corresponds to `B' mode start. A2 and A1 were taken as the power in the bin corresponding to the peak of each component.  Due to the shape of the profile, A2/A1 is sensitive to the selection of the bin for A2.  Errors were therefore calculated by shifting the bin for A2 by 1 in each direction and recalculating A2/A1.}
\label{fig:a2a1}
\end{center}
\end{figure}

While we do not have polarimetry for this observation and cannot directly confirm the previously reported changes in polarization, we can investigate the profile shape changes.  Figure~\ref{fig:a2a1} shows a stack of average profiles for the reconstructed `B' mode [Fig.~\ref{fig:p3hat_v_time_b}] along with the ratio of the amplitudes of the trailing component to the leading component (A2/A1) as a function of pulse number.  Due to strong RFI and the lower sensitivity of GMRT compared to AO, we are unable to measure A2/A1 as accurately as in \textit{Paper IV}, however we confirm that the trailing component is relatively brighter at `B' mode onset and grows relatively weaker as time progresses.

By fitting a function of the form $a+b \exp(-t/\tau)$ to A2/A1 we find a characteristic time $\tau = 1.7 \pm 1.0 hrs$ which agrees within the error to the time reported in \textit{Paper IV}.  The fit is not well constrained, and there is significant error on our measurements, however we can confirm the general picture presented by \textit{Paper IV} that the leading component gradually dominates the profile with a timescale on the order of one hour.

We can therefore describe the evolution of the modes in B0943+10.  At `B' mode onset, the average profile shape changes abruptly, a precursor component virtually disappears [\cite{Backus}], and a strong primary modulation feature becomes visible in LRF's.  Over the span of some 7.5 hr the frequency of the feature changes as an exponential function of the form of eq.(\ref{eq:f3obs}) by some 5\%.  The profile shape changes exponentially with the same time scale as $f_{3,obs}$ (\textit{Paper IV}).  The fractional linear polarization increases throughout the `B' mode (\textit{Paper V}).  At `B'-mode cessation the profile switches abruptly and the strong modulation feature disappears.  The PC brightens to detectable levels and the pulsar then spends around 2.2 hr in the `Q' mode.  Remarkably this behavior is a confirmation of the assorted observations reported in \cite{Backus} and \textit{Papers~I--V}.

\section*{IV. Change in modulation under carousel model}
\label{sec:IV}

We proceed to interpret the evolution of the primary modulation under the CM.  The analysis presented in \textit{Paper~I} and \cite{Backus} showed that the `B' mode of B0943+10 displays a rotating-subbeam carousel comprised.  For three PS's of several hundred pulses, spanning the entire `B'-mode, the number of subbeams was measured to be 20.  $f_{3,obs}$ is argued to be a first order alias of the true primary modulation frequency $f_{3,true}$ such that $f_{3,true}=1-f_{3,obs}$.  We can therefore find the true $P_3$ and the carousel circulation time ($\P3hat$ here, often referred to as $P_4$):
	\begin{equation}
P_3 = \dfrac{1}{1-f_{3,obs}}
	\label{eq:P3}
	\end{equation}
	\begin{equation}
\P3hat = 20 P_3 = \dfrac{20}{1-f_{3,obs}}
	\label{eq:P3hat}
	\end{equation}
In this case, $P_3$ and $\P3hat$ will evolve similarly to $f_{3,obs}$: they will increase as an exponential function of the form of eq.(\ref{eq:f3obs}), changing by about 5\% over the `B' mode.  Any fractional changes in $P_3$ will be reflected in similar changes in $f_{3,obs}$.

We may now investigate what changes in the carousel could account for the $\sim$5\% change in $P_3$.  For B0943+10, $P_3$ results from drifting subpulses.  Under the CM, the subbeam carousel rotates about the magnetic axis, causing subbeams to appear at different longitudes for different pulses.  This effect causes the observed subpulse drifting.

$P_3$ is therefore a measure of the time between adjacent subbeams crossing the same longitude in the profile.  Two variations can then account for a change in $P_3$: a change in the angular separation between subbeams (as measured in the plane perpendicular to the magnetic axis) or a change in the angular velocity of the carousel (hereafter: $\omega _c$ with a corresponding change in $\P3hat$.  In either case, small changes in the viewing geometry will not affect $P_3$.

It is not simple to resolve which of these effects causes the observed change in $P_3$, but to understand the physics of moding and pulsar emission we must answer this question.  We consider what observational evidence we have concerning this matter.

\subsection*{1. Change in carousel angular velocity $\omega_c$}
\begin{figure}
\begin{center}
\includegraphics[width=80mm]{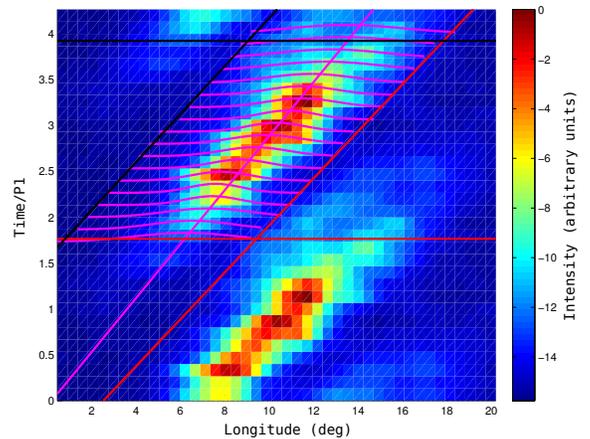}
\caption{Sample results of a modfold: the track of a subpulse across the pulse window for the MJD 53491 AO observation, performed over pulses 1-256.  Longitude in deg. is plotted on the horizontal axis and time in units of $P_1$ is plotted on the vertical axis.  A 256-pulse LRF was performed to determine $P_{3,obs}$.  The same pulse sequence was folded at $P_{3,obs}$ and plotted twice.  A bounding box (red and black) was drawn around the subpulse track.  Gaussians (magenta) were fitted to each row in the bounding box and a line (magenta) was then fitted to the peaks of the Gaussians, from which $m$ (see~eq. \ref{eq:slope}) was determined.}
\label{fig:slope_fit}
\end{center}
\end{figure}

If $\omega_c$ decreases, then the subbeams will cross a given longitude on the polar cap less frequently, thus changing the observed $P_3$.  Similarly, the rate of change of the position on the polar cap of a given subbeam will decrease.  This will cause the subbeam to appear to drift less quickly across the pulse window.

The drift rate of B0943+10 is fast enough that an individual subpulse cannot often be observed in consecutive pulses: after being observed in one pulse, the subpulse has left the pulse window by the next pulse.  To observe the track of a subpulse across the pulse window, we follow the method used in \textit{Paper~I}.  An LRF is performed on a 256-pulse segment of the pulse sequence from which the frequency of the primary modulation feature is determined.  The 256 pulses are then folded according to the phase position within $P_{3,obs}$.  This results in a single sequence of length $P_{3,obs}$ from which we can resolve the rate at which a subbeam crosses the pulse window.  To mitigate wrapping effects, the resulting sequence is plotted twice.  This process shall be referred to as a modfold.  Fig.~\ref{fig:slope_fit} shows the results of a modfold.

If $\omega_c$ alone determines the variation in $P_3$, $\P3hat$ must change by some 5\%, and there will be a corresponding change in the slope of the subpulse traverse (see Fig.~\ref{fig:slope_fit}).  Under the CM, we expect the traverse of a subbeam to deviate from a line due to our curved sightline (as projected on the polar cap) and the curvature of the carousel.  In the case of B0943+10, our sightline traverse is tangential: essentially cutting the edge of the cone, so across the $\sim$20$\deg$ pulse window we expect any deviation from linearity to be minimal, as observed.

We may therefore treat the subpulse traverse as linear.  Since we are concerned only with changes in the subpulse traverse, and since the shape of the traverse is determined by our viewing geometetry which is assumed to be constant, the linear approximation is further justified.  As such, the slope ($m$) is then determined by $P_{3,obs}$ and the spacing between adjacent subpulses in the pulse window ($P_2$):
\begin{equation}
m = \dfrac{P_{3,obs}}{P_2}
\label{eq:slope}
\end{equation}
For a constant $P_2$, a 5\% change in $P_{3,obs}$ will produce a 5\% change in $m$.

Fig.~\ref{fig:slope_fit} demonstrates the determination of $m$ for one 256-pulse subsequence.  A modfold is performed.  A bounding box is drawn around the track of a subpulse.  The subpulses are quite Gaussian in form, so a Gaussian is fitted to each row of intensities within the bounding box.  We then have a stack of Gaussians whose peaks align with the centers of the subpulses.  The peak positions of the Gaussians are fitted to a line, and the slope of this line is taken to be $m$.  The slopes so determined are fairly stable with respect to changes in the bounding box.

A 5\% change in $m$ is very small given the accuracy of the method described above.  The GMRT observation is not of sufficiently high quality to resolve $m$ clearly, so the older AO observations were used to determine $m$ as a function of time.  For the observations used below (see \textit{Papers I, IV and V}), $m$ was determined for the first 256-pulse subsequence in the `B' mode, then it was determined for a 256-pulse subsequence beginning 128 pulses later.  Then the process was reiterated until the end of the mode (or observation).

\begin{figure}
\begin{center}
\includegraphics[width=85mm]{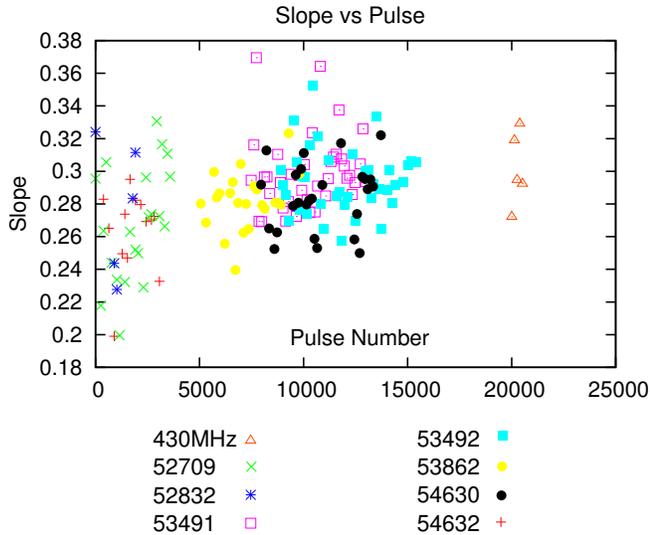} 
\caption{Drift-band slope in periods/$\deg$longitude as a function of time since `B'-mode onset for eight AO observations---as in Fig.~\ref{fig:slope_fit}.  The position within each `B'-mode observation was determined by measuring $f_{3,obs}$ and fitting to eq.(\ref{eq:f3obs}).  There appears to be a gradual increase in the slope ($m$) of about 5\%.}
\label{fig:slope_v_time}
\end{center}
\end{figure}

Fig.~\ref{fig:slope_v_time} shows $m$ as a function of time.  The $m$ values exhibit a great deal of scatter compared to the expected 5\% change we are trying to detect.  Errors in the measurement of $m$ would be expected to be of comparable magnitude: the extension of the subpulses is large compared to the total drift, their shapes deviate significantly from a Gaussian, the signal-to-noise ratio even after folding 256 pulses is still low enough to affect the fits, and the changing $P_3$ at the beginning of the `B' mode smears the apparent subpulse track.  

$P_{3,obs}$ decreases by about 5\% during the course of a `B' mode.  If changes in $\omega_c$ were solely responsible for the changes in $P_{3,obs}$, then $P_2$ would remain constant, and $m$ would vary proportionally to $P_{3,obs}$.  However, a 5\% decrease in $m$ is not observed---in fact, $m$ appears to increase by about 5\% (Fig.~\ref{fig:slope_v_time}).  Thus, a carousel velocity change alone cannot account for the observations.

\subsection*{2. Change in subbeam spacing}
We now explore the question of possible subbeam-spacing changes within the carousel.  If the number of subbeams $N$ is constant, then any such $P_2$ changes will occur on timescales small with respect to $\P3hat$.  Such changes would correspondingly affect $P_3$, but if $m$ (eq.\ref{eq:slope}) and $\omega_c$ remain constant, then $P_3$ will remain constant on average.  $\P3hat\approx$ 37$ P_1$, so in a given 256-pulse subsequence the carousel completes about 7 rotations.  Thus $P_3$ determined by an LRF performed over the subsequence will not be significantly affected.

A change in $P_3$ must therefore be attributed to a change in the number of subbeams: this will affect the average angular spacing of the subbeams.  According to the current model of B0943+10, the number of subbeams is $\sim$20 (\textit{Paper I}).  The smallest change in the average subbeam spacing must therefore be 5\%, and in turn $P_3$ will change by 5\%.  This conflicts what is observed: the primary modulation frequency appears to change smoothly by 5\% over the course of a `B'-mode apparition (see Fig.~\ref{fig:p3hat_v_time})---not in discrete jumps.

However, if the number of subbeams changes on timescales less than 256 pulses, smooth changes in the observed $P_3$ could be explained.  If the carousel has multiple configurations each with a different number of sparks, then the fraction of time it spends in each configuration will determine the observed $P_3$.  Gradual changes in the fractional time spent in each configuration would then cause gradual changes in $P_3$.


In order to distinguish variations in the subbeam spacing we can determine $P_2$.  As with the determination of $m$, a $P_2$ change of around 5\% could explain the observed $P_3$ variations, but achieving such precision is difficult.  As discussed in \textit{Paper~I}, $P_2$ is around 10$\deg$ of longitude.  This is large with respect to the profile width, so it is difficult to observe single pulses with two subpulses: few pulses exhibit more than one and, in the few that do so, they fall on the profile wings where they are characteristically faint.  Thus, we cannot measure $P_2$ directly.

\begin{figure}
\begin{center}
\includegraphics[width=80mm]{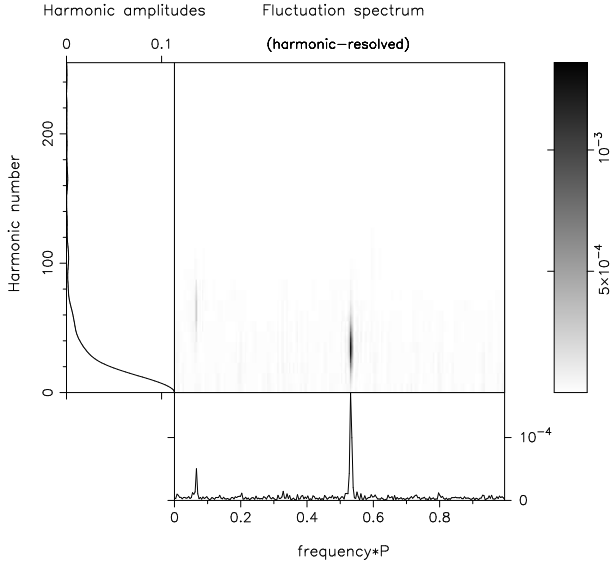}
\caption{Harmonically-Resolved Fluctuation spectrum (HRF) performed on the first 256 pulses of the MJD 53492 AO observation.  The frequency spectrum folded at $f_3$ is plotted in the bottom panel, excluding the bins corresponding to the pulsar frequency and its harmonics.  The intensity of the pulsar harmonics as a function of harmonic number is plotted in the left panel.  The center panel shows the intensity of the features lying at $(N+f)/P_1$ as a function of pulsar harmonic number $N$ and frequency $f$ where $0<f<\frac{1}{P_1}$.}
\label{fig:hrf}
\end{center}
\end{figure}

\begin{figure}
\begin{center}
\includegraphics[width=80mm]{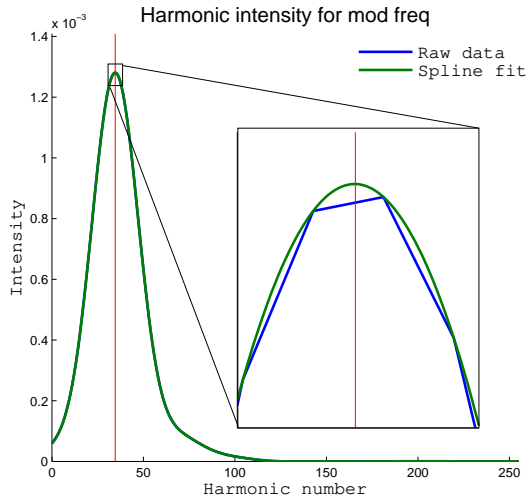}
\caption{A plot of modulation intensity {\it vs.} harmonic number for primary modulation feature, using the HRF computed in Fig.~\ref{fig:hrf}.}
\label{fig:peak_harmonic}
\end{center}
\end{figure}

\begin{figure}
\begin{center}
\includegraphics[height=80mm,angle=-90]{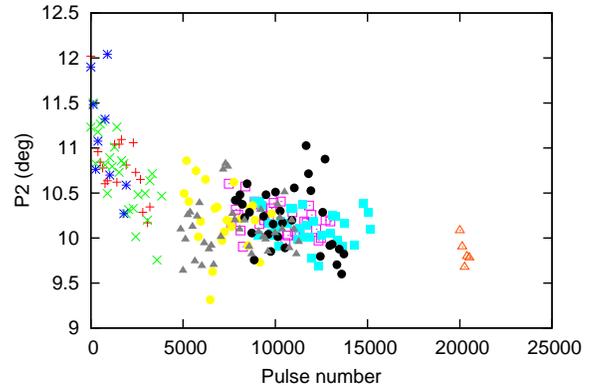}
\caption{$P_2$ as a function of time in the `B' mode for nine AO observations.  $P_2$ was measured by performing an HRF as in Fig.~\ref{fig:hrf}.  Using the frequency bin corresponding to the primary fluctuation feature, the peak of the features lying at $(N+f_3)/P_1$ was found, from which $P_2=360\deg/N_{max}$.}
\label{fig:p2_v_time}
\end{center}
\end{figure}

We therefore follow the analysis presented in \textit{Paper~I}.  A Fourier transform was performed on an unfolded time series.  The frequency of the pulsar and its harmonics are strongly visible in the frequency domain.  Additionally, components due to $P_3$ falling at $(N+f_3)/P_1$ are visible, where $N$ is the harmonic number of the pulsar rotation frequency.  The amplitudes of these components peak around $N_{max}\approx$ 35 (see Figs~\ref{fig:hrf} and \ref{fig:peak_harmonic}). $N_{max}$ then represents the harmonic of $P_1$ associated with the primary subpulse-modulation frequency---or, in analogy with eq.(\ref{eq:slope}), $f_{3,obs}=1/(N_{max}P_2)$.  $P_2$ (given its units in degrees) can then be calculated as $360\deg/N_{max}$, where $N_{max}$ is the harmonic number corresponding to the peak amplitude.

To determine $f_3$, the fluctuation spectrum is folded at the pulsar frequency.  Fig.~\ref{fig:hrf} shows an harmonically resolved fluction spectrum (HRF).  The folded frequency domain is plotted in the bottom panel with the pulsar harmonics (now the DC component) removed.  The left panel shows the amplitude of the pulsar harmonics as a function of harmonic number.  The center panel shows the amplitude of the components of the folded Fourier transform as a function of the pulsar harmonic number.

The primary modulation peak in the folded Fourier domain determines $f_3$.  Fig.~\ref{fig:peak_harmonic} shows the amplitudes of the $(N+f_3)/P_1$ components for the HRF plotted in Fig.~\ref{fig:hrf}.  A cubic spline fit was performed to find the position of the peak.  Generally, the amplitudes change smoothly, so that the splines fit the amplitudes well.  There is typically a clear maximum.

In this manner, $P_2$ was determined as a function of position in `B' mode for the available AO observations.  The position of each AO observation in `B' mode was found by exploiting the changes in $f_{3,obs}$.  For each observation, $f_{3,obs}$ was found as a function of time by performing 256-pulse LRFs.  The results were then fitted to eq.(\ref{eq:f3obs}) and are plotted in Fig.~\ref{fig:p2_v_time}.  $P_2$ appears to decrease in a manner similar to the increase in $P_3$, changing by about 10\% over the course of the `B' mode.  

\begin{figure}
\begin{center}
\includegraphics[height=80mm]{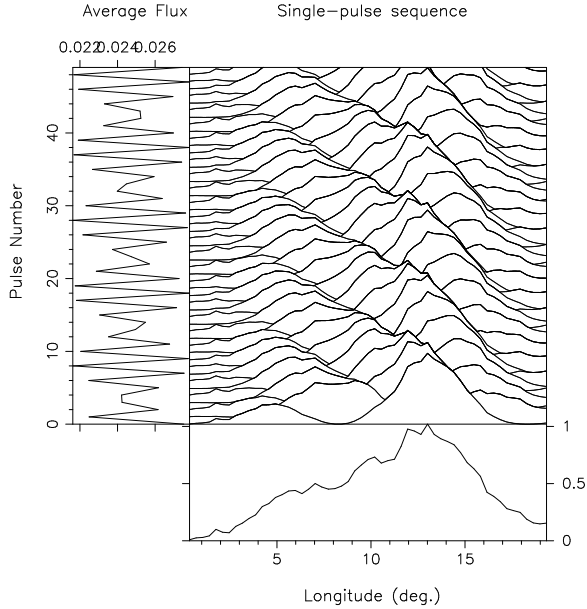}
\caption{A simulated pulse sequence for B0943+10.  A 50-pulse stack is plotted in the center panel.  The average of those 50 pulses is plotted in the bottom panel, and the flux for each pulse is plotted in the left panel.  Subpulses were modeled by a positively valued sine wave.  Pulses were modeled by multiplying the subbeam sine wave by an average profile which changed from pulse to pulse.  For each pulse, the subpulse phase was increased by $2\pi P_1/P_3$ to simulate subpulse drifting.}
\label{fig:simulated_ps}
\end{center}
\end{figure}

The average pulse profile changes shape with the same time scale as the changes in $P_3$ and with a manner very similar to the observed changes in $P_2$.  We were concerned that the changing profile shape or the changing $P_3$ could affect our measurement of $P_2$, thereby producing the illusion of a change in $P_2$.  To resolve this question, a modeled pulse sequences was produced to test our method of finding $P_2$ (see Fig.~\ref{fig:simulated_ps}).

The subpulses were modeled as a positively valued sine wave (\ie, $1+sin\phi$).  The period of the sine wave was set to be 10$\deg$ longitude corresponding to the observed $P_2$.  The pulses were modeled by multiplying the sine wave by a `B' mode average profile taken from an AO observation with the off-pulse window set to 0.  The phase of the subpulse sine wave was increased by $2\pi P_1/P_3$ from pulse to pulse to simulate subpulse drifting.

To check if changes in the average profile affect measurements of $P_2$ two methods were followed.  First, 256-pulse average profiles were calculated from an AO observation beginning at `B' mode onset.  The pulse window over which the profile was calculated was incremented by one pulse and the process was completed for the remainder of the `B' mode.  In order to mimic the pulse shape changes, the pulses were modeled using these average profiles according to the method above.  For the second method, a 256-pulse profile was calculated for the beginning and the end of a `B' mode.  An exponential function was used to interpolate the changes in each bin of the profile for the several thousand pulses in the `B'  mode, thus producing a smoothly changing average profile to be used to model pulses.

The modeled pulse sequence was then analyzed in the same way as the observations to determine $P_2$.  No change in $P_2$ greater than a fraction of a percent was observed using the modeled data, from which we conclude that the gradual changes in profile shape cannot account for the observed change in $P_2$.  

To see if the change in $P_3$ affects measurements of $P_2$ we used both a constant and a changing `B'-mode average profile while altering $P_3$ as a function of pulse number according to eq.(\ref{eq:f3obs}) given the relation in eq.(\ref{eq:P3}).  No change was detected in the measurement of $P_2$ in the modeled sequence, from which we conclude that the changing $P_3$ cannot account for the observed changes in $P_2$.

Thus, we observe a decrease in $P_2$ by about 10\% during the `B' mode.  This change appears to be fairly continuous.  $P_{3,obs}$ decreases by about 5\% over a similar timescale.\footnote{$P_3$ variations can in principle ``Doppler shift'' $P_2$; \eg, see \cite{Gupta2004}, eq. (5).  However, for B0943+10 here these effects are at the 1\% level.}  By eq.(\ref{eq:slope}) we would expect these two effects to produce a $\sim$5\% increase in the slope of the subpulse track, which agrees with the measurements of $m$ (Fig.~\ref{fig:slope_v_time}).  We conclude that there is an intrinsic change in the slope of the subpulse track.

Under the CM, this change in the slope indicates a change in $\omega_c$.  Accounting for aliasing, the observed decrease in $P_{3,obs}$ corresponds to an increase in the true $P_3$.  From eqs.\ref{eq:f3obs} and \ref{eq:P3} we see there is an increase in $P_3$ of 4.1\% during the `B'-mode.  Due to our tangential sight-line traverse, $P_2$ should be inversely proportional to the number of subbeams $N$.  According to the CM, $\P3hat = 2\pi / \omega _c = N P_3$.  Thus we can say:
\begin{equation}
P_2 \propto \frac{1}{N}
\label{eq:P2_v_N}
\end{equation}
\begin{equation}
P_3 \propto \frac{1}{\omega _c N}
\label{eq:P3_v_N}
\end{equation}
Combining these two equations tells us:
\begin{equation}
\omega _c \propto \frac{P_2}{P_3}
\label{eq:wc_P2_P3}
\end{equation}
Given a 10\% decrease in $P_2$ and a 4.1\% increase in $P_3$ during the `B'-mode, eq.(\ref{eq:wc_P2_P3}) tells us there should be a 13.5\% decrease in $\omega _c$.  Since $\P3hat \propto 1/\omega _c$, there is a 16\% increase in $\P3hat$ during the `B'-mode.


\section*{V. Discussion}
\label{sec:V}
We have presented a long, 8-hr observation of B0943+10 using the GMRT telescope in India.  Such a long observation of the pulsar is important for resolving questions about the behavior of the long modes in B0943+10.  We confirm the gradual evolution of the primary fluctuation feature which corresponds to subpulse drifting in the `B' mode.  From `B' mode onset until `B' mode cessation, $f_{3,obs}$ varies according to eq.(\ref{eq:f3obs}) with a characteristic time of $\tau = 1.2$ hr.  This supports the conclusions of earlier papers in this series that during the course of a `B' mode there are changes in the average profile with a similar characteristic time as well as a gradual increase in the fractional linear polarization during a `B' mode.

As a rough estimate, we find that `B' modes will persist for around 7.5~hrs in length and `Q' modes some 2.2~hrs.  Significant deviation from these numbers is expected (and is observed for the `Q' mode).

The previous papers in this series have argued that the emission from B0943+10 can be understood according to the subbeam-carousel model, wherein 20 evenly spaced subbeams rotate around the magnetic axis in about 37~$P_1$.

We detect a decrease in $P_2$ of about 10\% during the course of a `B' mode.  Under the carousel model, this indicates an increase in the number of subbeams.  Given about 20 subbeams, there should be two more subbeams at the end of `B' mode than at the beginning.  \cite{Backus} found three occasions, lasting some several hundred pulses, where the number of subbeams was measurable.  These three occasions spanned the entire `B'-mode.  Using the analysis presented in \S~IV, $P_2$ is found to decrease as expected for these occasions.  For all three, $N$ was found to be 20, which seems to contradict the fact that $P_2$ measurements indicate an increase in $N$.  However, in \cite{Backus}, the measurement of $N$ at the beginning of the `B'-mode has a large error due to low signal-to-noise in the LRF used to compute $N$.  Thus the findings of \cite{Backus} are commensurate with a carousel composed of 20 subbeams at the end of the `B'-mode and fewer subbeams (perhaps 18) at the beginning.\footnote{See \cite{Backus} figures 1, 2, and 3 for the measurements of $N$.}

For $N \approx 20$, one would expect discrete jumps of about $1/20 = 5\%$ in our measurements of $P_2$.  However this is not observed.  Rather, the changes in $P_2$ appear continuous, albeit with significant scatter.  Our measurements of $P_2$ are calculated over several hundred pulses, so if the carousel model is correct the pulsar must switch between two configurations with a different number of subbeams on timescales much less than several hundred pulses.  This indicates that the pulsar preferentially spends more time in one configuration at the beginning of `B' mode and gradually spends more time in other configurations.

By measuring $P_2$, $P_3$, and the slope of subbeams as they cross the pulse window---which gives a measure of how quickly a given subbeam passes through a degree of longitude---we find that over the course of a `B' mode $\P3hat$ increases by some 16\% rather than the 5\% found in \textit{Paper~I}.

We may now speculate on a possible physical explanation of the gradual changes in $\P3hat$.  According to the partially screened gap model (PSG), the ${\bf E}\times{\bf B}$ drift velocity of the subbeams around the magnetic axis should be related to the temperature of neutron star surface at the polar cap.  In the PSG model, the vacuum gap as proposed by \cite{RS1975} is actually partially filled with ions (or possibly electrons).  Heating of the neutron star surface due to bombardment by relativistic particles causes the ejection of charge carriers into the vacuum gap.  This shields the potential drop $\Delta V$ across the vacuum by some factor $0 < \eta < 1$ such that $\Delta V = \eta \Delta V_{RS}$, where $\Delta V_{RS}$ is the potential drop across a purely vacuum gap \citep{Gil2003}.

The drift velocity, and hence $\omega _c$, should be proportional to $\Delta V$, thus $\P3hat \propto \eta ^{-1}$ \citep{Gil2006}.  We observe a 16\% increase in $\P3hat$ over a `B'-mode apparition, so we can say:
\begin{equation}
\frac{\P3hat,_2}{\P3hat,_1} = \frac{\eta _1}{\eta _2}=1.16
\label{eq:T1}
\end{equation}
According to \cite{Gil2003}:
\begin{equation}
\eta = 1 - exp[C(1-T_c/T_s)]
\label{eq:T2}
\end{equation}
where $T_s$ is the surface temperature of the neutron star, $T_c$ is the critical temperature at which the potential drop is completely shielded (i.e. $\Delta V = 0$) and $C$ is a constant.  As argued by \cite{Gil2003}, $T_s$ should be at most a few percent less than $T_c$, so the argument of the exponent in eq.(\ref{eq:T2}) is close to 0.  Performing a Taylor expansion and keeping the terms up to first order:
\begin{equation}
\eta \approx -C(1-T_c/T_s)
\label{eq:T3}
\end{equation}
From eq.(\ref{eq:T1}) and eq.(\ref{eq:T3}) we find that the ratio of the surface temperature at the end of a `B'-mode to the temperature at the beginning should be:
\begin{equation}
\frac{T_{s,2}}{T_{s,1}} \approx \frac{1.16 T_c}{T_c + 0.16 T_{s,1}}
\label{eq:T4}
\end{equation}
Now, $T_s \lesssim T_c$.  So for $0.9 T_c < T_s < T_c$ we find that $1 < T_{s,2}/T_{s,1} < 1.014$.  This means that according to the PSG model we should expect to see an increase in the surface temperature of less than 1.4\% over the course of a `B'-mode.  B0943+10 has been observed in the X-ray regime \citep{Zhang2005}, and it may be possible with long simultaneous X-ray and radio observations to detect such small changes in the temperature.  This would be strong evidence in favor of the PSG model, along with the carousel model.

\vspace{12 pt}
\noindent {\bf Acknowledgments:}
We are pleased to acknowledge Janusz Gil, Svetlana Suleymanova, and Geoffrey Wright 
for their critical readings of the manuscript.  We thank the GMRT staff for support during the observation.  Two of us (IB and JMR) thank the 
National Centre for Radio Astrophysics in Pune for their generous hospitality.  
Portions of this work were carried out with support from US National Science 
Foundation Grants AST 99-87654 and 08-07691.  Arecibo Observatory is 
operated by Cornell University under contract to the US NSF.  This work used 
the NASA ADS system.

\bibliographystyle{mn2e}
\bibliography{mybib}{}
\end{document}